\documentstyle[aps,epsf,twocolumn,prl]{revtex}

\begin{document}
\draft

\twocolumn[\hsize\textwidth\columnwidth\hsize\csname @twocolumnfalse\endcsname
\title{The Lower Critical Dimension of the $XY$ Spin Glass}

\author{J. Maucourt and D. R.  Grempel}

\address{D\'epartement de Recherche Fondamentale sur
la Mati\`ere Condens\'ee \\
SPSMS, CENG, 17, rue des Martyrs, 38054 Grenoble
Cedex 9, France}

\date{\today}
\maketitle
\widetext
\begin{abstract}
\noindent

We investigate the $XY$ spin-glass model in two and three dimensions 
using the 
domain-wall renormalization-group method. The results for systems 
of linear sizes up to  $L=12$ (2D) and $L=8$ (3D) strongly suggest 
that the lower critical dimension for spin-glass ordering may be 
$d_{\rm 
c}\approx 3$ rather than four as is 
commonly believed. Our 3D data favor the scenario of a low but 
finite spin-glass 
ordering temperature below the chiral transition but they are also 
compatible with the system 
being at or slightly below its lower critical 
dimension.
\end{abstract}


]
\narrowtext

 It has been known since the 
early work of Villain\cite{villain} that  
frustrated planar models possess, in addition 
to the continuous degeneracy associated with 
spin rotations, a discrete two-fold degeneracy 
associated 
with the invariance of the Hamiltonian under a reflection about an 
arbitrary direction. As a 
consequence, each plaquette 
has an Ising-like degree of 
freedom, the chirality,  that corresponds to 
the handedness of the configuration of the spins 
around it.  
It is also well established\cite{kawa-tane,ray-moore,kawa1,kawa2} 
that in two and 
three  
dimensions chiral and spin variables 
decouple at long distances and order separately\cite{jain}.
In 2D long range chiral and spin order appear simultaneously  
at zero temperature, but the transition is unusual in that there 
are two 
independent divergent correlation lengths as $T\to 
0$\cite{kawa-tane,ray-moore}. While  
there is convincing evidence\cite{kawa-tane,kawa1,kawa2} that in 3D 
the chiral-glass 
transition occurs at
a finite temperature $T_{\rm CG}$,  existing Monte Carlo 
simulations\cite{kawa2,jain-young} and studies of the scaling of 
defect energies at zero temperature\cite{kawa-tane,kawa2} suggest 
that spin-glass 
order sets in at $T_{\rm SG}=0$ just as in two dimensions. These, as 
well as older numerical results\cite{banavar,mac,morris} have 
led to the belief that the 
lower critical dimension (LCD) for spin-glass order in this model is  
$d_{c}\ge 4$, a conjecture that is widely accepted even if a rigorous 
proof has 
turned out to be elusive\cite{nish-oz,schw-young,oz-nish}. 
The limitations inherent to
the numerical methods  raise some doubts about the robustness of this 
conclusion, 
however. The systems studied with the 
defect-energy 
method are rather small. While the  
domain wall energy flows towards weak coupling with 
increasing linear size $L$ for systems with $L\le 4$\cite{morris}, 
more recent  
simulations\cite{kawa2}  
for $L\le 6$ show a tendency towards saturation for the 
bigger sizes. It is thus conceivable that the systems studied up to 
now are 
still far from the scaling regime. The Monte Carlo 
simulations\cite{kawa2,jain-young} were for much larger lattices
 but the rapid increase in the  
thermalization times with decreasing $T$ makes it impossible to
 attain the relevant temperature region, $T\ll T_{\rm CG}$. 
It appears that is also difficult to reach the asymptotic regime  
with this technique. Indeed, Kawamura\cite{kawa2} has noticed that 
the Binder 
function associated with the spin-glass order parameter fails to  
show scaling behavior in the temperature range covered by his 
simulations, $T\gtrsim T_{\rm CG}$. The spin-glass 
susceptibility does seem to scale but Jain and Young\cite{jain-young} 
found that their 
data can be fitted with 
comparable accuracy by assuming such different values of the 
spin-glass 
critical temperature as $T_{\rm SG}=0$ and $T_{\rm SG}=0.45 J$. 
In view of such uncertainties we may still regard the nature 
of the low-temperature phase of the 3D $XY$ spin-glass as an open 
problem.

In this paper we study the $XY$ spin-glass model in two and three 
dimensions with the domain-wall renormalization-group method 
(DWRG)\cite{banavar,mac} . 
Using a new and powerful algorithm for the search of ground-state 
energies we have been able to study 
systems with linear 
sizes up to $L=12$ (in 2D) and $L=8$ (in 3D). In both cases our 
largest system contains more than twice as many spins as have been  
considered before. Our results in 2D agree with those found by other 
authors\cite{kawa-tane}, confirming that the scaling regime had been 
reached in previous simulations. This turns out {\em not} to be the 
case in three dimensions, where we find that there is a crossover 
between small- and large-$L$ behavior at $L\approx 5$. The 
domain-wall energies scale as $W\sim L^{\lambda}$.  The sign of the 
stiffness exponent $\lambda$ may be positive or negative depending on 
whether the system is above or below its LCD. We find $\lambda_{\rm 
c}=0.5\pm 0.2$ 
and $\lambda_{\rm s}=0.056\pm 0.11$ for chiral and spin domain walls, 
respectively. The fact that $\lambda_{\rm c}>0$ is the signature that 
there is long-range chiral order at low-temperature in the system as 
found  by 
other   
authors\cite{kawa-tane,kawa2}. The smallness of $\left|\lambda_{\rm 
s}\right|$ constitutes to our knowledge the first numerical evidence 
that the LCD of the 
$(\pm J)$ $XY$ model may be close to three. Two scenarios are 
compatible 
with the measured value of $\lambda_{\rm s}$, i) a 
spin-glass 
transition at a finite temperature  $T_{\rm SG}\ll T_{\rm CG}$ or  ii) 
a zero-temperature 
transition with the correlation length $\xi \sim T^{-\nu}$ 
and $\nu\gg 1$. Statistics favors the former.

The Hamiltonian of the model is 
\begin{eqnarray}
H = -\sum_{<ij>} J_{ij}\ \cos \left(\theta_{i}-\theta_{j}\right),  
\label{hamil}
\end{eqnarray}
where the sum runs over all pairs of nearest-neighbor sites of the 
2D square or 3D cubic lattices. The exchange couplings $J_{ij}$ are 
random independent 
variables that take the values $J=+1$ and $J=-1$ with equal 
probability. In the DWRG method\cite{banavar,mac}  one studies the 
sensitivity 
of the system to changes in the boundary conditions at $T=0$. The 
ground-state energies of an ensemble of systems of size $L^D$ are 
calculated using periodic (P) and anti-periodic (AP) boundary 
conditions along a given direction while keeping fixed boundary 
conditions 
along the $(d-1)$ remaining directions. The width of the distribution 
of differences of ground-states energies, $W_{\rm s}(L)=[\left(E_{\rm 
P}-E_{\rm A}\right)^2]^{1/2}_{J}$ is interpreted as an effective 
coupling constant between blocks of $L^D$ spins. This is expected 
to scale as $W(L)\sim L^{\lambda}$ for large enough 
$L$. If the stiffness exponent, $\lambda$, is positive the rigidity 
of 
a block 
diverges with its size, a sign that there is long-range order in the 
system. If $\lambda<0$ there exists a length scale 
$L^{-|\lambda|}\sim 
T$ beyond which the effective coupling between blocks becomes smaller 
than the temperature. This scale  is identified with the correlation 
length and the correlation-length exponent is obtained from 
$\nu=1/|\lambda|$\cite{morris}. Chiral ordering may be studied 
similarly except 
that the calculation requires knowledge of the ground-state energy
 for reflective boundary conditions as well as the other 
two\cite{kawa-tane}. 

The success of this approach  relies on the availability of an 
efficient and accurate way of finding the lowest-lying states of the 
system. In the usual spin-quench algorithm\cite{walker} (SQA) one 
randomly 
generates long sequences of 
metastable configurations among which one 
hopes to find the ground-state or states  sufficiently close in 
energy. The number of metastable states of a frustrated 
system increases rapidly with its size (probably 
exponentially\cite{morris}) and so does  
the number of trials that need to be performed to have a 
non-negligible chance of finding a relevant state. This fact 
restricts severely the sizes of 
the systems that can be studied using this method. 
In order to go beyond the limits of the SQA we have developed a 
ground-state search algorithm of far greater efficiency. This 
algorithm is inspired by some of the morphological features of the 
low-lying states of the frustrated $XY$ model that are as 
follows\cite{gawiec,ga-gr-ma}. 
i) In a given sample there are regions with a low density of 
frustrated plaquettes where the spins form almost collinear domains, 
and 
regions where frustration is 
high and the spin arrangement looks random. The position, size and 
shape of the domains are mostly determined by the random bond 
realization and hardly vary from one state to another. ii) Apart from 
smooth spin-wave-like distortions, the essential differences between 
the spin configurations of any two low-energy states are large 
amplitude, {\em  almost rigid} rotations of the individual domains, 
and chirality reversals of a fraction of the frustrated plaquettes in 
the regions separating them. The energies of states that do not have 
this structure are much higher\cite{ga-gr-ma,gawiec}.
Our method suceeds in preserving the domain structure at each stage 
of the procedure by treating the spins in 
high local fields (domain spins) differently from those in the more 
frustrated regions. In this way the appearance of uninteresting 
high-energy configurations becomes unlikely. The algorithm will be 
sketched in the following paragraph. Full details will be given in a 
separate publication\cite{ga-gr-ma}.   

The initial (or parent) state of the sequence, $\{\theta_{\rm 0}\}$, 
is obtained by a  
conjugate-gradient minimization (CGM) of the energy  with a random 
spin  distribution as initial condition. Next, new spin 
configurations are generated  by iterating the following loop any 
number of times.
i) The spins in the $n$-th configuration $\{\theta_{\rm n}\}$ are 
divided into two groups according to whether their local field 
$h_i=\sum_j J_{ij}\cos\left(\theta_i-\theta_j\right)$ is greater or 
smaller than an  
appropriately chosen threshold field, $0\le h_{t}\le 2D$ (see below). 
The spins in the first group are {\em defined} as the domain 
spins. 
ii) Existing correlations between the domains and the rest of the 
system 
are broken by means of a random rigid   
rotation of the former.
iii) A fraction $p$ of the spins 
in weak local fields are randomly picked and their orientations are 
randomly reset. 
iv) The energy of 
the subsystem of domain spins is minimized {\em with the rest of the 
spins 
held fixed}. 
v) The resulting configuration is let to relax performing a CGM of 
the {\em total} energy of the system. The outcome is the next state 
in the sequence, $\{\theta_{\rm n+1}\}$.
vi) The energy of this state is stored and $h_{t}$ is 
rescaled if necessary (see below). 
vii) End of the loop.

\begin{figure}
\epsfxsize=3.5in
\epsffile{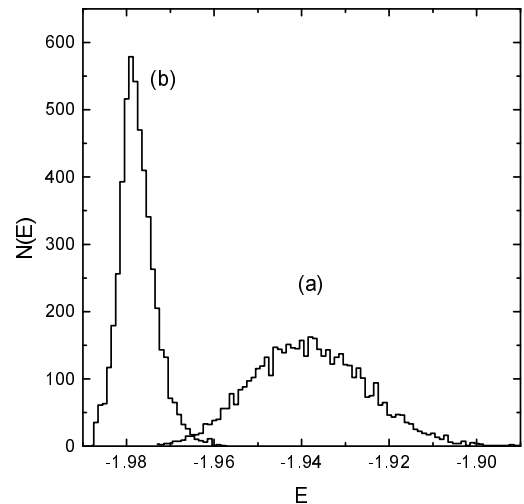}
\caption{Distribution of the energy per spin E of two series of 5 000 metastable states 
each of one realization of the $XY$ $J=\pm 1$ spin glass on a 512-site 
3D lattice. The states have been generated using the spin-quench 
method  
(a) and the algorithm described in the text (b).
}
\label{fig1}
\end{figure}

The parameter $h_{t}$ must be carefully chosen for this algorithm to 
be efficient. It defines the degree of collinearity required of an 
ensemble of spins for it to qualify as a domain ($h_{t}=2D$ for 
perfectly aligned spins in an homogeneous environment). 
If $h_{t}$ is set very high, too 
many spins are involved in step iii) above and the domain structure 
is  
 not preserved. In this regime the method is equivalent to the SQA 
where {\em all} the spins are randomly reset at each step.
On the other hand,  too small a threshold field leads to trapping, 
{\it i.e.}, all the states of the sequence are in the vicinity 
of the parent state. There is an 
optimum value of the threshold field in between these 
extreme cases. In the running version of our code 
the performance of the algorithm is continuously monitored and a 
procedure has been devised that allows 
$h_{t}$  to self-adjust when degradation is 
detected.  This method is particularly appropriate for the treatment 
of large systems. This is illustrated in 
Fig.\ \ref{fig1} which shows the energy distributions of two 
sequences of   
metastable states of the same realization of a disordered 3D system 
of  
$N=8^{3}$ spins with anti-periodic boundary conditions. 
Histograms (a) and (b) have been obtained from 5000 energy levels 
each generated using the SQA and our algorithm, respectively. 
In the latter case we have verified that trapping had not 
occurred by making sure that the same histogram results   
from sequences originating from  different parent states. 
Distributions (a) and (b) are quite different. Whereas the histogram 
of the energy levels obtained with the 
standard method is wide and peaks at high energies, that of our 
sequences is  much narrower and is concentrated in the 
low-energy end of the spectrum.  It is remarkable that {\em most} of 
the 
energies found with the new method lie in 
a region where the SQA did not find {\em any} state after the 
same  number of trials. The calculation with our method takes only 
20{\%} more CPU time than that with the traditional one.

\begin{figure}
\epsfxsize=3.5in
\epsffile{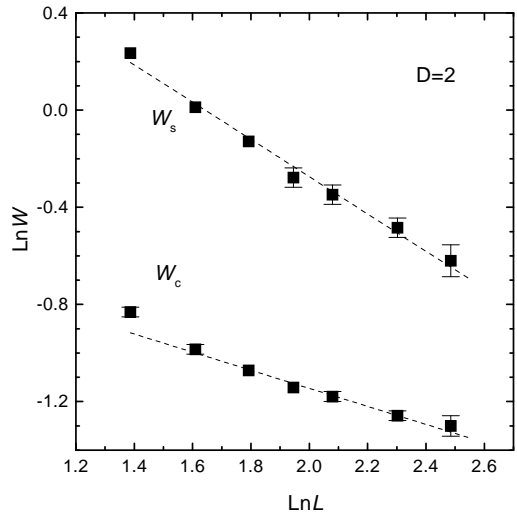}
\caption{The $L$-dependence of the spin (squares) and 
chiral (circles) domain-wall energies of the $(\pm J)$ $XY$ spin-glass 
model on $L\times L$ square lattices. Where not drawn 
the statistical error bars are smaller than the size of the data points. The 
dashed lines are the power-law fits mentioned in the text.
}
\label{fig2}
\end{figure}

We have used this method to perform DWRG calculations for two- and 
three-dimensional systems. Much is  known about the  
two-dimensional case which therefore serves as a test of our 
methods.  
We have determined spin and chiral defect energies 
for systems of size 
$L\times L$ with  $L$=4,5,6,7,8,10 and 12. 
The disorder averages were taken over 128000 
($L$=4), 64000 ($L$=5), 12800 ($L$=6 and 7), 6400 ($L$=8), 2560 
($L$=10) and 1280 ($L$=12) independent 
bond configurations, respectively. The ground state energy was 
estimated   
from the analysis of sequences containing 20 ($L$=4), 50 ($L$=5), 100 
($L$=6), 200 
($L$=7 and 8), 1500 ($L$=10), and 3000 ($L$=12) states. Low-energy 
states were  
 accepted as ground-state candidates only if they {\em and} 
their chirality-reversed partners appeared several times in the 
sequence at widely spaced positions. This guarantees that the states 
in the sequence come from well separated regions of phase-space. A 
log-log plot of the size dependence of the two-dimensional defect 
energies is 
shown in Fig.\ \ref{fig2}.  The symbols are 
the numerical results and the dashed 
lines are fits to a power-law. The fits are of good quality even for 
small $L$. The renormalized stiffness decreases with increasing $L$ 
for both types 
of domain wall indicating that spin and chiral variables only order 
at zero temperature. From the slopes determined  by the fits 
we can 
compute the correlation length exponents for the two transitions. We 
find $\nu_{\rm s}=1.29\pm 
0.02$ and $\nu_{\rm c}=2.57\pm 0.003$ for the spin and chiral 
order parameters, respectively. These  
values are in  good agreement with the results of 
Kawamura and Tanemura\cite{kawa-tane} who find $\nu_{\rm s}=1.2\pm 
0.15$ 
and $\nu_{\rm c}=2.6\pm 0.3$, respectively in  their DWRG 
calculations for systems with $L\le 8$. Moreover, the value of the 
chiral 
exponent deduced from our data is very close to the correlation 
length 
exponent of the 2D Ising spin-glass obtained 
by Monte Carlo\cite{bhatt} ($2.6\pm 
0.4$) or transfer-matrix\cite{cheung} ($2.59\pm 0.13$) methods. This 
agrees with the view that the chiral transition in the $XY$ spin 
glass model is in the Ising universality class.

\begin{figure}
\epsfxsize=3.5in
\epsffile{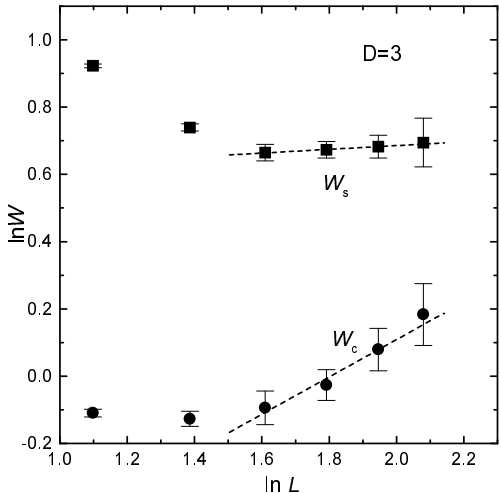}
\caption{The $L$-dependence of the spin (squares) and 
chiral (circles) domain-wall energies for the $(\pm J)$ $XY$ spin-glass 
model on $L\times 
L\times L$ simple cubic lattices. The 
dashed lines are the power-law fits mentioned in the text.
}
\label{fig3}
\end{figure}

We now turn to the results for three-dimensional systems. Much longer 
sequences of 
stationary states are needed to determine the ground-state energy in 
this case. Our series consist of 100 ($L$=3), 
200 ($L$=4), 500 ($L$=5), 1000 ($L$=6), 3000 ($L$=7) and 5000 ($L$=8) 
states, respectively. Moreover, the results for the bigger lattices  
were further checked by repeating the calculation starting form three 
different parent states. Sample averages were taken over  25600 
($L$=3), 6400 ($L$=4), 
1280 ($L$=5,6), 640 ($L$=7) and 128 ($L$=8) samples, respectively. 
The $L$-dependence of the effective couplings in three dimensions is
 shown by the symbols in Fig.\  \ref{fig3}. An important difference 
with the 2D case is that a crossover region is clearly seen around 
$L$=5. Power-law behavior, shown in the figure by the dashed lines, 
is only observed for the largest lattices. We first discuss the 
chiral degrees of freedom. As found in previous 
studies\cite{kawa-tane,kawa2}, $W_{\rm c}$ flows towards 
strong coupling signaling the existence of finite-temperature 
chiral-glass transition 
in the infinite system. From the fits shown in the figure  
we can estimate the chiral stiffness exponent $\lambda_{\rm 
c}=0.56\pm 0.18$.

The spin domain-wall-energy decreases rapidly with $L$ for small 
sizes but it exhibits a much slower variation for $L\gtrsim 5$. This 
behavior was not observed in older simulations\cite{kawa-tane} but 
Kawamura's more recent DWRG results  distinctively show the beginning 
of the 
 saturation of the spin defect energy 
 for $L\sim 5$. Fitting the results for the four largest sizes with a 
power-law
we find the spin-stiffnes exponent 
$\lambda_{\rm s}=0.056\pm 0.11$. The large error bar is due to poor 
statistics in the case of our largest size for which the sample 
average could only be taken over 128 configurations of the bonds. 
The error in the determination 
of the ground-state energy, estimated from a comparison of the 
results of searches conducted starting from different parent states, 
is much lower. It is worth mentioning that if we make the fit 
omitting the last point the result is $\lambda_{\rm s}=0.052\pm 
0.03$. 

The smallness of $|\lambda|$ suggests that the LCD 
of the $(\pm J)$ $XY$ model is close to three. The data statistically 
favor $\lambda_{\rm s}\gtrsim 0$. If this is the case the spin-glass 
transition temperature should be finite with $T_{\rm SG}\ll  T_{\rm CG}$ 
as implied by our finding that $\lambda_{\rm c}\gg |\lambda_{\rm 
s}|$. The numerical results are also compatible with a second and far 
less exciting possibility, a zero-temperature transition with an unusually 
large correlation-length exponent.

There exist no theoretical objections against a spin-glass 
transition {\em below} $T_{\rm CG}$ since the 
proofs\cite{nish-oz,oz-nish} of the absence of SG ordering in the 
three-dimensional model at finite $T$ fail if   
reflection-symmetry is broken\cite{kawa2,schw-young,oz-nish} as is the 
case when chiral-glass order is present. The possibility of a finite-temperature SG transition discussed here is thus restricted to the case of the 
$XY$ model for which one can convincingly argue that $T_{\rm CG}\ne 0$. 
The case of the Heisenberg model seems quite different in that the role of the chiral variables is not as clear as in the 
planar model and there is no evidence of the existence of chiral order
at finite temperatures. Therefore we still expect isotropic three-dimensional 
spin-glass models to have an ordered phase only at zero temperature\cite{banavar,mac,matsu}.

We thank Professors H. Nishimori and H. Kawamura for helpful  
correspondence and T. Ziman for a critical reading of the manuscript. 
The calculations reported in this work have been 
performed on the 256-processor CRAY T3E parallel computer at the 
`Centre Grenoblois de Calcul Vectoriel'. We thank the staff for 
technical help.

\end{document}